\documentclass[12pt]{iopart}

\usepackage{amsfonts}
\DeclareFontFamily{OT1}{rsfs}{}
\DeclareFontShape{OT1}{rsfs}{m}{n}{<5> rsfs5 <7> rsfs7 <10> rsfs10
}{}
\DeclareSymbolFont{mathrsfs}{OT1}{rsfs}{m}{n}
\DeclareSymbolFontAlphabet{\mathrsfs}{mathrsfs}
\newcommand{\R}{\mathrsfs{R}}

\newcommand*{\Cet}[1]{|{#1}\rangle}
\newcommand*{\Bra}[1]{\langle{#1}}

\newcommand{\Up}{\uparrow}
\newcommand{\Dw}{\downarrow}

\newcommand{\UpCet}{\Cet{\Up}}
\newcommand{\DwCet}{\Cet{\Dw}}

\newcommand{\UpUp}{\Bra{\Up}\Cet{\Up}}
\newcommand{\UpDw}{\Bra{\Up}\Cet{\Dw}}
\newcommand{\DwUp}{\Bra{\Dw}\Cet{\Up}}
\newcommand{\DwDw}{\Bra{\Dw}\Cet{\Dw}}

\newcommand{\CUp}{C^{\Up}}
\newcommand{\CDw}{C^{\Dw}}

\newcommand*{\en}[1]{\Big|{#1}\Big|^2\ln\Big|{#1}\Big|^2}
\newtheorem{axiom}{Axiom}

\begin{document}

\title[The information theory and the collapse of a wavefunction]{The information theory and the collapse of a wavefunction at the measurement of a spin $1/2$ projection}

\author{Denys Bondar}

\address{Department of Physics and Astronomy, University of Waterloo, 200 University Avenue West, N2L 3G1 Waterloo, Ontario, Canada}

\ead{\mailto{kdf\_studio@yahoo.com}}

\begin{abstract}
From the point of view of the information theory, a model of the collapse phenomena at the measurement of a spin $1/2$ projection is developed. This model phenomenologically includes an observer. The model allows not only to determine the state of a system after the measurement but also to compute the state of the observer. The state of the observer is equivalent to the operator of a spin projection which the observer will measure at the next measurement.
\end{abstract}
\submitto{\JPA}

\section{Introduction}

In this article, we are going to treat the collapse of a wavefunction phenomena by means of the information theory. The collapse phenomena is one of the most challenging task in modern theoretical physics and nowadays it is an object of intense studies because of a number of applications. Here are some references about the measurement theory and the collapse phenomena \cite{Ruseckas, Pearle1, Kent, Habib, Pearle2, Bassi, Nakagomi}. The role of an observer in a measurement process has been discussed in  \cite{Nakagomi, Menskii, Mensky, Green, Brillouin2, Brillouin1, Neumann}. In addition to its original purpose -- communication, the information theory has been successfully applied not only to physics \cite{Green, Brillouin2, Brillouin1, Nalewajski} but also to biology and evolution \cite{Avery}. And as it can be seen that the employment of the information theory to a problem allow us to see the matter from a different point of view.

Based on ideas from \cite{Menskii, Mensky, Neumann, Brillouin1}, an axiomatic model of the collapse phenomena will be formulated. We will consider the simplest case of the collapse -- the collapse after the measurement of a spin $1/2$ projection. Before we start, it is convenient to describe the spin projection from the point of view of non relativistic quantum mechanics.

The operator of a spin projection on an axis, that is determined by
the normalized vector
\begin{equation}
\overrightarrow{n} = (\sin\theta\cos\varphi,
\sin\theta\sin\varphi, \cos\theta), \quad 0\leq\theta\leq\pi,
\quad 0\leq\varphi < 2\pi,
\end{equation}
has the following form
\begin{equation}
\sigma(\theta,\varphi) = \sigma_n =
\overrightarrow{\sigma}\cdot\overrightarrow{n} = \left(
\begin{array}{cc}
\cos\theta & \sin\theta e^{-i\varphi} \\
\sin\theta e^{i\varphi} & -\cos\theta
\end{array}
\right),
\end{equation}
where $\overrightarrow{\sigma}=(\sigma_x,\sigma_y,\sigma_z)$ is
the Pauli vector operator with components
$$
\sigma_x = \left(\begin{array}{cc}0 & 1\\1 &
0\end{array}\right),\qquad \sigma_y = \left(\begin{array}{cc}0 &
-i\\i & 0\end{array}\right),\qquad \sigma_z =
\left(\begin{array}{cc}1 & 0\\0 & -1\end{array}\right).
$$
The eigenvalues and the orthonormal eigenvectors of the operator
$\sigma_n$ are the following
\begin{eqnarray}
\sigma_n\UpCet = +1\UpCet, && \sigma_n\DwCet = -1\DwCet,\\
\UpCet = \left(
\begin{array}{c}
\cos(\theta/2) e^{-i\varphi} \nonumber\\
\sin(\theta/2)
\end{array}\right), \quad&&
\DwCet = \left(
\begin{array}{c}
-\sin(\theta/2)e^{-i\varphi} \label{EginF}\\
\cos(\theta/2)
\end{array}\right).
\end{eqnarray}

\section{The first and second axioms}

Before developing the model, it is useful to describe the phenomena in the simplest terms. The scheme of the collapse of a wavefunction is pictured on the first figure. Time goes from the left side to the right side of the figure. The vertical bold line shows the time moment of the collapse which is also the time of the measurement. As it can be seen on the picture, our quantum system is characterized by the wavefunction $\Cet{\Psi}$ and an observer wants to measure the projection $\sigma(\theta_i, \varphi_i)\ $ before the collapse. According to the postulates of quantum mechanics, the system is either $\UpCet_i$ or $\DwCet_i$ after the measurement. As it will be understood later, it is convenient to introduce an unknown projection $\sigma(\theta_f, \varphi_f)$ ($\UpCet_f$ and $\DwCet_f$ are its eigenvectors) which will be interpreted like\textit{ the next projection that the observer will want to measure at the next measurement.}

\begin{figure}
\caption{The scheme of the collapse for the measurement of a spin
projection}
\begin{center}
\begin{picture}(300,80)
\put(150,40){\oval(100,80)}
\put(100,40){\llap{$\Cet{\Psi}=\CUp_i\UpCet_i +\CDw_i\DwCet_i$}}
\put(200,40){ \ $\UpCet_i$ \ or \ $\DwCet_i$}
\put(150,60){\llap{$\sigma(\theta_i, \varphi_i)\ $}}
\put(145,20){\llap{$\UpCet_i$, $\DwCet_i$}} \put(150,60){$\
\sigma(\theta_f, \varphi_f)$} \put(153,20){$\UpCet_f$, $\DwCet_f$}
\put(0,35){\thicklines\vector(1,0){100}}
\put(200,35){\thicklines\vector(1,0){100}}
\put(150,0){\linethickness{5pt}\line(0,1){80}}
\end{picture}
\end{center}
\end{figure}
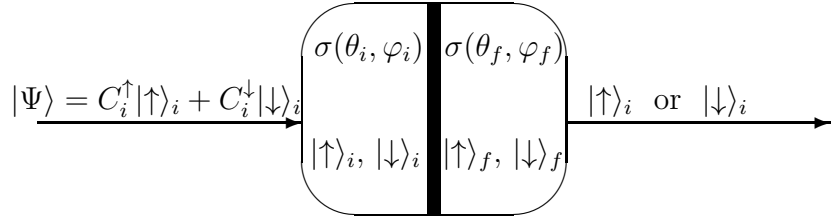

The following expansions take place 
\begin{eqnarray}
\Cet{\Psi} = {}_i\Bra{\Up}\Cet{\Psi}\UpCet_i + {}_i\Bra{\Dw}\Cet{\Psi}\DwCet_i, && 
\Cet{\Psi}={}_f\Bra{\Up}\Cet{\Psi}\UpCet_f + {}_f\Bra{\Dw}\Cet{\Psi}\DwCet_f, \\
\UpCet_i = {}_f\UpUp_i\UpCet_f + {}_f\DwUp_i\DwCet_f, \quad
&& \DwCet_i = {}_f\UpDw_i\UpCet_f + {}_f\DwDw_i\DwCet_f.\label{UpDwExp}
\end{eqnarray}
Entropies which connected to the previous expansions are presented by
\begin{eqnarray}
S_i &=& -\en{{}_i\Bra{\Up}\Cet{\Psi}} - \en{{}_i\Bra{\Dw}\Cet{\Psi}},\\
S_f &=& -\en{{}_f\Bra{\Up}\Cet{\Psi}} - \en{{}_f\Bra{\Dw}\Cet{\Psi}},\\
S_{\Up} &=& -\en{{}_f\UpUp_i} - \en{{}_f\DwUp_i}, \label{SUp}\\
S_{\Dw} &=& -\en{{}_f\UpDw_i} - \en{{}_f\DwDw_i}. \label{SDown}
\end{eqnarray}
Before the measurement, the initial entropy of the quantum system is $S_i$; the final entropy of the system is equal to zero, because the entropy of a pure state is zero. But according to the negentropy principle of information (\cite{Brillouin1}, p~153) (the conservation of the sum of entropy and information), our initial entropy has been transformed into (`bound') information which the observer has obtained about the physical system after the measurement. However, \textit{let us represent the state of the observer by the projection of the spin, which the observer wants to measure.} Therefore, the state of the observer after the measurement is characterized by $\sigma(\theta_f, \varphi_f)$. So in our case the amount of this information is represented by $S_f$. Now we are able to formulate the first axiom
\begin{axiom}
\begin{equation}\label{Ax1}
S_i(\theta_i, \varphi_i)=S_f(\theta_f, \varphi_f).
\end{equation}
\end{axiom}
It should be pointed out that the left part of equation (\ref{Ax1}) is a known value ($\theta_i$ and $\varphi_i$ are fixed), but the right part of equation (\ref{Ax1}) is unknown, because the parameters $\theta_f, \varphi_f$ have not been determined. We will discuss how to calculate these angles later. 

However, there is one exception, when the observer wants to measure the projection of the spin on the same axis, that the initial state is an eigenstate of the operator of that projection, i. e. the wavefunction of the system is either $\Cet{\Psi}=\UpCet_i$ or $\Cet{\Psi}=\DwCet_i$. In this case equations (\ref{Ax1}) is a trivial equality. According to the postulates of quantum mechanics, the initial state is also the final state, i. e. there is no collapse in this case. Hence, the state of the observer cannot be changed which implies that $\theta_f=\theta_i$ and $\varphi_f=\varphi_i$. We will exclude this trivial case in further investigations. 

We will return to the computation of the parameters $\theta_f$ and $\varphi_f$. Primarily, it should be pointed out that the following relation directly goes from equation (\ref{UpDwExp})
\begin{equation}
\Big|{}_f\UpUp_i\Big|^2 + \Big|{}_f\UpDw_i\Big|^2 = 1.
\end{equation}
Taking into account equations (\ref{SUp}) and (\ref{SDown}), we can conclude that
\begin{equation}\label{SEquiv}
S_{\Up} \equiv S_{\Dw}.
\end{equation}

Now we know only equation (\ref{Ax1}) that connects the values of two angels $\theta_f$ and $\varphi_f$; but this is not enough to determine their values. Fortunately, there is one additional clue from the information point of view, under the name generalized Carnot's principle (\cite{Brillouin1}, p~153). We will use generalized Carnot's principle in the form of equation (12.5 a)  (\cite{Brillouin1}, p~154), that indicates the decreasing of the amount of information that an observer obtains about a system. In thermodynamics, we have the principle of the increasing of an entropy for an isolated system and it is well known that the equilibrium state of the isolated system is one of maximum entropy. Using generalized Carnot's principle and following the ideology of thermodynamics, we can make a statement that the equilibrium state of the  system is one of minimum information. The first axiom has fixed the amount of information that the observer obtains about our system in the current measurement, so we shall minimize information that the observer will obtain in the next measurement. 

Let us consider the next measurement. For this measurement, the initial entropy of the system is either $S_{\Up}$ or $S_{\Dw}$. However, according to equation (\ref{SEquiv}), there is no difference between these two entropies and we will consider $S_{\Up}$ only. According to the first axiom,  $S_{\Up}$ is the amount of information that the observer will receive. Finally, we reach the second axiom:
\begin{axiom}
$$
S_{\Up}(\theta_i, \varphi_i, \theta_f, \varphi_f) \longrightarrow min.
$$
\end{axiom}
Now we are able to define the value of the parameters $\theta_f$ and $\varphi_f$ like the minimum of $S_{\Up}$ with condition (\ref{Ax1}).

Let us rewrite equation (\ref{Ax1}) in the terms of our parameters. We can assume without the lose of  generality that a normalized vector has the form
\begin{equation}
\Cet{\Psi} = \left(\begin{array}{c} \sqrt{\rho}e^{-i\tau}\\
\sqrt{1-\rho} \end{array}\right), \qquad 0\leq\rho\leq 1, \qquad
0\leq\tau\leq2\pi.
\end{equation}
From equation (\ref{EginF}) we obtain 
\begin{equation}
\Big| {}_f\Bra{\Up}\Cet{\Psi} \Big|^2 =
\rho\cos^2\frac{\theta_f}2 + (1-\rho)\sin^2\frac{\theta_f}2 +
\sqrt{\rho(1-\rho)}\sin\theta_f \cos(\varphi_f-\tau).
\end{equation}

But we still cannot answer the main question: ``What is the final state of the system after the collapse?'' Unfortunately, there is no additional hints in the information theory. This means that the information theory can only particularly describe the collapse phenomena. In order to answer the main question, we must look at the phenomena from another point of view. We will try to do this in the following section.

\section{The third axiom}

Discussing the principle of the psycho-physical parallelism, Von Neumann has mentioned
\begin{quotation}
``But in any case, no matter how far we calculated – to the mercury vessel, to the scale of the thermometer, to the retina, or into the brain, at some time we must say: and this is perceived by the observer. That is, we must always divide the world into two parts, the one being the observed system, the other the observer. [...] Experience only makes statements of this type: an observer has made a certain (subjective) observation; and never any like this: a physical quantity has a certain value.'' (\cite{Neumann}, p~419-420)
\end{quotation} 
The first two axioms phenomenologically represent the `observed' system, but to have the complete description of the collapse phenomena, we have to include the `observer.' 

We shall follow the ideas by Menskii \cite{Menskii}. First of all, we should accept that 
\begin{quotation}
``The ability of a human (and of any living creature) referred to as consciousness is the same phenomenon
as that which is termed the reduction of state or alternative selection in the quantum theory of measurement and which appears in Everett's concept as the separation of the single quantum world into classical alternatives.'' (\cite{Menskii}, p~401)
\end{quotation}
After this definition, it is natural to ask the question: \begin{quote} ``What set of alternatives [...] is preferred among all possible sets from the viewpoint of living creatures?'' (\cite{Menskii}, p~402) \end{quote} According to Menskii, the answer is the following
\begin{quotation}
``Each alternative [...] describes the behavior of a microscopic system under measurement and [...] also represents all its (macroscopic) environment, i.e., the whole world. This picture of the world emerges in the consciousness of a living creature. When the world in this picture behaves in accordance with classical laws, it is `locally predictable.' [...] Seeing the predictable world around, a living creature can work out the optimal strategy for survival in this world.'' (\cite{Menskii}, p~403)
\end{quotation}
This answer can also be formulated like ``classical alternatives: prerequisite to the existence of life'' (\cite{Menskii}, p~402).

We will try to formalize this principle. There are at least two methods on how to do this. The first and probably the most complete one is to use the same mathematical models that successfully have been used in economics for modeling the following situation: how a person can make an optimal choice among available ones (see for example \cite{Hirshleifer}). However, to apply these methods to the collapse phenomena, the uniqueness of this case must be well understood. Unfortunately, we do not have this knowledge yet. Therefore, in order to have the complete model, we shall employ a rough method.

Following the ideology of the principle of least action, we introduce a real function $\R$ which will act as an action in our approach. \textit{Let us name this function as the function of risk, because the value of the function will be interpreted as the measure of the risk of an alternative} (the larger the value of the function is, the more dangerous the alternative is for a living creature). 

As far as the form of the function $\R$ is concerned, it is obvious that this function is a function of the parameters $\theta_f$, $\varphi_f$, and the spin projection $s (s=\pm1)$, i. e. $\R=\R(\theta_f, \varphi_f; s)$. We can formulate the third axiom.
\begin{axiom}
The parameters $\theta_f$, $\varphi_f$, and $s$ satisfy the condition $\R(\theta_f, \varphi_f; s) \to min$.
\end{axiom}
According to the symmetry of the problem, the function $\R$ has to be periodic: $\R(\theta_f+\pi, \varphi_f+2\pi; s)=\R(\theta_f, \varphi_f; s)$. 

The main problem is to determine the manifest from of the function $\R$ and we will do this in the next work.

\section{Conclusion}

We have built the axiomatic model of the collapse of a spin $1/2$ projection. This model can be generalized not only to high values of a spin but also to weak measurements.

Summarizing, we provide the scheme which describes how to use these axioms for determining the state of the system after the collapse:
\begin{itemize}
 \item To find the value of the parameters $\theta_f$ and $\varphi_f$ as the solution of the extremal problem for $S_{\Up}$ with restricion (\ref{Ax1}),
 \item To find the value of the spin projection $s$ which minimize the function of risk $\R$.
\end{itemize}
Finally, not only have we computed the final state of our system, but also the value of the parameters $\theta_f$ and $\varphi_f$ which represent the projection that the observer will measure at the next measurement. 

\ack
The author wants to thank Dr. Robert Lompay for very important discussions.

\end{document}